# On mathematical theory of selection: Discrete-time population dynamics

Georgy P. Karev
NCBI, NIH
3600 Rockville Pike, Bethesda, MD 20894, USA
karev@ncbi.nlm.nih.gov

## 1. Introduction and background

According to the Darwinian theory, natural selection uses the genetic variation in a population to produce individuals that are adapted to their environment. The Fisher's Fundamental theorem of natural selection (*FTNS*) states [Fisher 1999]: "the rate of increase in fitness of any organism at any time is equal to its genetic variance in fitness at that time, except as affected by mutation, migration, change of environment and the effect of random sampling".

Many versions of the FTNS were proved within the frameworks of different exact mathematical models (see equations (2.5), (2.6) below) but the actual biological content of the theorem, even if it were true mathematically, had been a subject of discussion in the literature for decades ([Ewens 1969], [Frank & Slatkin 1992], [Edwards 1994], etc.). Fisher himself noted that the FTNS only holds subject to important assumptions. A disappointment in the standard FTNS was expressed by G. Price [Price 1972]: "A grave defect is the matter of the shifting standard of "fitness" that gives the paradox of [the average fitness] tending always to increase and yet staying generally close to zero".

Price [Price 1970, 1972] derived an equation to describe any form of abstract selection (see equation (2.3)). The FTNS and the so-called covariance equations (see equation (2.4)) are its particular cases. Price claimed that his equation is an exact, complete description of evolutionary change under all conditions [Price 1972a], in contrast to the FTNS and covariance equation, where the "environment" is fixed. The Price equation was applied not only to biological problems, such as evolutionary genetics sex ratio, kin selection (see [Rice 2006, ch.6], [Crow 1976], [Price 1972], [Hamilton 1970], etc.), but also to social evolution [Frank 1998], evolutionary economics [Knudsen 2004], etc.

Lewontin seems to be the first who noticed that the Price equation can not be used alone as a propagator of the dynamics of a trait forward in time and hence it is in no case a complete description of evolutionary change [Lewontin 1974]. The Price equation, being a mathematical *identity* is not dynamically sufficient, i.e., it does not allow one to predict changes in the mean of a trait beyond the immediate response (if only the value of covariance of the trait and fitness at this moment is known).

However, the asymptotical behavior of selection systems is known in the following sense: if a limit distribution exists and is asymptotically stable, then it is concentrated in a finite number of points, which are the points of global maximum of an average reproduction coefficient on a support of the initial distribution. This "extremal principle" proved in [Semevsky, Semenov 1982], [Gorban 2005] is a generalization of the Haldane principle [Haldane 1990]. So, based on the known results, the behavior of systems under selection can be predicted at the first time step and "at infinity".

Here, we develop a theory of general selection systems with discrete time and explore the evolution of selection systems at the entire time interval where a global solution of the system is defined. We prove that the distribution of the system can be explicitly determined and computed at any time, so all statistical characteristics of interest, such as the mean values of the fitness or any trait can be computed effectively. In particular, we show that the problem of dynamic insufficiency for the Price equations and for the FTNS can be resolved within the framework of selection systems if the entire initial distribution of the parameters is known.

## 2. Discrete-time models. Inhomogeneous maps as mathematical models for selection

Let us assume that a population consists of individuals, each of which is characterized by its own vector-parameter $\mathbf{a}=(a_1,\ldots a_k)$. Here, we do not specify the vector-parameter $\mathbf{a}$, whose components may be arbitrary traits, e.g., $a_i$ could be the number of alleles of $i$-th gene, as in simple genetic models. We prefer to think of $\mathbf{a}$ as the entire genome or as a set of specific genes. For the general case, we will denote $\{\mathbf{a}\}=A$.

Let $l(t,\mathbf{a})$ be the population density (informally, the number of individuals with a given vector-parameter $\mathbf{a}$) at moment $t$. In general, the fitness of an individual depends on the individual vector-parameter $\mathbf{a}=(a_1,a_2,\ldots a_n)$ and on the "environment" that depends on time. Then in the next time instant

$$l(t+1,\mathbf{a})= w_t(\mathbf{a})\, l(t,\mathbf{a}). \tag{2.1}$$

where the reproduction rate $w(t,\mathbf{a})$ (fitness, by definition) is a non-negative function.

Let $N(t) = \int_A l(t,\mathbf{a})\,d\mathbf{a}$ be the total population size,

$$P_t(\mathbf{a})= l(t,\mathbf{a})/N(t) \tag{2.2}$$

be the current probability density function (pdf). It is important to note that, if $P_t(\mathbf{a}^*)=0$ for particular $\mathbf{a}^*$ at some instant $t$, then $P_{t'}(\mathbf{a}^*)=0$ forever, for all $t'>t$. Hence, selection system (2.1) describes the evolution of a distribution with a support that does not increase over time. Any (measurable) function $\varphi_t(\mathbf{a})$ can be considered as a random variable over the probabilistic space $(A,P_t)$; we will denote $E_t[\varphi_t]=\int_A \varphi_t(\mathbf{a})P_t(\mathbf{a})\,d\mathbf{a}$. Then

$N(t+1)=N(t)E_t[w]$ and $P_{t+1}(\mathbf{a})= P_t(\mathbf{a})w_t(\mathbf{a})/E_t[w_t]$.

So, $E_{t+1}[\varphi_{t+1}] =E_t[\varphi_{t+1}w_t]/ E_t[w]$ for any r.v. $\varphi_t(\mathbf{a})$.

Next, for any sequence $\{s_t, t=0,1,\ldots\}$, denote $\Delta_t s= s_{t+1}- s_t$. Let $z_t(\mathbf{a})$ be a character of an individual with the given vector-parameter $\mathbf{a}$, which can vary with time. Then

$E_t[w_t]\, \Delta_t E_t[z_t] = E_t[w_t](E_{t+1}[\Delta_t z] + E_{t+1}[z_t] - E_t[z_t]) = E_t[w_t \Delta_t z] + E_t[z_t\, w_t] - E_t[w_t]E_t[z_t] =$
$\mathrm{Cov}_t [z_t\, w_t] + E_t[w_t \Delta_t z_t]$.

We obtained the Price equation within the framework of the general model (2.1):

$$\Delta_t E_t[z_t] =(\,\mathrm{Cov}_t[z_t w_t] + E_t[w_t \Delta_t z])/ E_t[w_t]. \tag{2.3}$$

We see that the Price equation holds for any particular fitness and any character; actually, it is a mathematical identity under the model (2.1)-(2.2), so it is impossible to "solve" it, i.e., to compute the temporal dynamics of a particular character beyond the immediate response without additional information or suppositions. Note that, if the character $z$ does not depend on $t$, i.e., $\Delta_t z=0$, then (2.3) implies the covariance equation ([Robertson 1968], [Li 1967], [Price 1970]):

$$\Delta E_t[z] =\mathrm{Cov}_t [w_t, z]/ E_t [w_t] . \tag{2.4}$$

If $z=w$ in equation (2.3), then

$$E_t[w_t]\, \Delta E_t [w_t] = \mathrm{Var}_t [w_t]+ E_t [w_t \Delta_t w]. \tag{2.5}$$

If the fitness does not depend on time, i.e. $\Delta_t w=0$, then

$$\Delta E_t[w]=\mathrm{Var}_t[w]/E_t[w], \tag{2.6}$$

which is the standard form of FTNS.

We will show that, for a large class of models (2.1)-(2.2), the current pdf of the parameter, $P_t(\mathbf{a})$, can be computed if we suppose that the initial pdf of the parameter, $P_0(\mathbf{a})$, is known in its entirety. Then, any term in (2.3) can be computed independently of others, hence the problem of dynamical insufficiency disappears.

This class of models is defined by a certain condition on the reproduction rate $w_t(\mathbf{a})$. For model (2.1) $w_t(\mathbf{a})>0$ and hence $w_t(\mathbf{a})=\exp(F_t(\mathbf{a}))$, where $F_t(\mathbf{a})=\ln[w_t(\mathbf{a})]$ is the "logarithmic reproduction rate". Taking into account that any smooth function $f_t(\mathbf{a})$ can be approximated by a finite sum of the form $\sum_i \varphi_i(\mathbf{a})g_i(t)$, where $\varphi_i$ depends on $\mathbf{a}$ only, and $g_i$ depends on $t$ only, we will suppose further that the fitness is of the form

$$w_t(\mathbf{a})=\exp[\sum_{i=1}^{n} \varphi_i(\mathbf{a})g_i(t)]. \tag{2.7}$$

Formula (2.7) defines the map from the set of all possible genotypes $\{\mathbf{a}\}=A$ to the set of corresponding fitnesses. Generally speaking, determination of this map is one of the central problems in biology. Within the framework of the master model (2.1), (2.7), we consider an individual fitness dependently on a given finite set of traits labeled by $i=1,\ldots n$. The function $\varphi_i(\mathbf{a})$ describes the quantitative contribution of a particular $i$-th trait (or gene) to the total fitness, and $g_i(t)$ describes a possible variation of this contribution with time depending on the environment, population size, etc. Let us emphasize that we do not suppose that contributions of different traits are independent of each other.

### 3. The main statistical characteristics of a population and their evolution

The following main Theorem 1 shows that master model (2.1), (2.2), (2.7) can be reduced to a non-autonomous map and completely explored. Let us denote

$$K_t(\mathbf{a}) = w_0(\mathbf{a}) \dots w_t(\mathbf{a}) = \exp(\varphi_1(\mathbf{a})G_1(t) + \dots \varphi_n(\mathbf{a})G_n(t)), \qquad (3.1)$$

where $G_i(t) = g_i(0) + \dots + g_i(t)$. It is easy to see that $w_t K_{t-1} = K_t$ and $l(t+1, \mathbf{a}) = K_t(\mathbf{a}) l(0, \mathbf{a})$.

We could think of the function $K_t(\mathbf{a})$ as the reproduction coefficient for the $[0,t]$-period or, for short, $t$-fitness. Let us note, that sometimes the functions $g_i(t)$ and hence $G_i(t)$ can be well defined not for all $0 < t < \infty$, but only for $0 < t < T$, where $T$ is a certain finite time moment. Accordingly, all assertions below can be valid only for $t < T$. Below we do not specify this condition if it is not necessary.

Denote $\vartheta = (\varphi_1, \varphi_2, \dots \varphi_n)$ and let $p(t; \vartheta)$ be the pdf of the random vector $\vartheta$ at $t$ moment, i.e., $p(t; x_1, \dots x_n) = P_t(\varphi_1 = x_1, \dots \varphi_1 = x_1)$. The master model defines a complex transformation of the distribution $P_t(\mathbf{a})$ or, equivalently, the pdf $p(t; \vartheta)$. Let $\lambda = (\lambda_1, \dots \lambda_n)$; denote

$$M_t(\lambda) = \int_A \exp(\sum_{i=1}^n \lambda_i \varphi_i(\mathbf{a})) P(t; \mathbf{a}) d\mathbf{a} = \int_A \exp(\sum_{i=1}^n \lambda_i x_i) p(t; x_1, \dots x_n)) dx_1, \dots dx_n$$

the moment generation function (mgf) of the pdf $p(t; \vartheta)$ of the random vector $\vartheta$.

The initial pdf $p(0; \vartheta)$ is supposed to be given. The mgf of the initial distribution, $M_0(\lambda)$, is crucially important for the theory developed below. For example, $E_0[K_t] = M_0(\mathbf{G}(t))$ where we denoted $\mathbf{G}(t) = (G_1(t), \dots G_n(t))$.

**Theorem 1.** *Let $P_0(\mathbf{a})$ be the initial pdf of the vector-parameter $\mathbf{a}$ for inhomogeneous map* (2.1), (2.7). *Then*

1) *The population size $N_t$ satisfies the recurrence equation*

$$N_{t+1} = N_t E_t[w_t]; \qquad (3.2)$$

*and can be computed by the formula*

$$N_t = N_0 E_0[K_{t-1}] = N_0 M_0(\mathbf{G}(t-1)) \qquad (3.3)$$

2) *The current pdf $P_t(\mathbf{a})$ satisfies the recurrence equation*

$$P_{t+1}(\mathbf{a}) = P_t(\mathbf{a}) w_t(\mathbf{a}) / E_t[w_t] \qquad (3.4)$$

*and can be computed by the formula*

$$P_t(\mathbf{a}) = P_0(\mathbf{a}) K_{t-1}(\mathbf{a}) / E_0[K_{t-1}] \qquad (3.5)$$

The current form and evolution of the distribution are very important for applications; the complete description of $P_t(\mathbf{a})$ is given by Theorem 1. The following lemma is key for practical investigation of the evolution of different distributions.

**Lemma 1.** $M_t(\lambda) = M_0(\lambda + \mathbf{G}(t-1)) / M_0(\mathbf{G}(t-1))$.

One can easily show with the help of Lemma 1 that, if the r.v. $\varphi_i$ are independent at the initial instant, then they stay independent for any $t \in [0, T)$. The general case of correlated r.v. $\varphi_i$ is of major practical interest; it helps to explore the evolution of the joint distribution of the quantitative contributions of different traits $\varphi_1, \varphi_2, \dots \varphi_n$ to the total fitness and the dynamics of inhomogeneous population depending on correlations between the traits (biologically, it is the case of absence of epistasis).

**Definition**. A class $S$ of probability distributions of the random vector $\vartheta = (\varphi_1, \dots \varphi_n)$ is called invariant with respect to model (2.1), (2.7), if $p(0, \vartheta) \in S \Rightarrow p(t, \vartheta) \in S$ for all $t < T$.

Let $\mathbf{MS}$ be the class of moment-generating functions for distributions from the class $S$. The criterion of invariance immediately follows from Lemma 1.

**Invariance criterion**. A class $S$ of pdf is invariant with respect to model (2.1), (2.7) if and only if $M_0(\lambda) \in \mathbf{MS} \Rightarrow M_0(\lambda + \mathbf{G}(t)) / M_0(\mathbf{G}(t)) \in \mathbf{MS}$ for all $t < T$.

**Remark**. We can prove with the help of this criterion (similarly to [Karev 2005]) that many important distributions, such as multivariate normal, polynomial, Wishart', and natural exponential distributions, are invariant with respect to the master model.

### 4. Inhomogeneous models with density-dependent fitness

The theory developed in s.3 for population models with the fitness of form (2.7) can be applied only if the time-dependent components $g_i(t)$ are known explicitly. As a rule, it is not the case for most interesting and realistic models where the time-dependent component should be computed depending on the current characteristics of the population. For example, a well-known logistic-type model, which accounts for the "deterioration of environment" accompanying population growth, corresponds to the function $g(N_t)=1-N_t/B$ where $B$ is the caring capacity. Allee-type models accounting additionally for the low boundary $b$ of the population density correspond to the function $g(N_t)=(1-N_t/B)(N_t/b-1)$.

In general, the population regulation ("change of the environment") may depend not only on the total population size but also on so called "regulators" having the form of some averages over the population density:

$$S_i(t)= \int_A s_i(\mathbf{a})l(t,\mathbf{a})d\mathbf{a}, \quad H_i(t)= \int_A h_i(\mathbf{a})P(t,\mathbf{a})d\mathbf{a}$$

where $s_i(\mathbf{a}), h_i(\mathbf{a})$ are appropriate functions. For example, if $h(\mathbf{a})=s(\mathbf{a})$ is a biomass of the individual with parameter $\mathbf{a}$, then $H(t)$ is an average biomass, $S(t)$ is a total population biomass and the population growth rate may depend on $S(t)=N_t H(t)$. The total population size $N_t$ is also a regulator with $s(\mathbf{a})=1$.

So, let us specify the theory developed in s.3 for the case of model
$l(t+1,\mathbf{a}) = l(t,\mathbf{a}) w_t(\mathbf{a})$,

$$w_t(\mathbf{a})=\exp[\sum_{i=1}^n u_i(S_i(t))\varphi_i(\mathbf{a})+\sum_{j=1}^m v_j(H_j(t))\psi_j(\mathbf{a})] \qquad (4.1)$$

where the individual fitness $w_t(\mathbf{a})$ may depend on some regulators. The main new problem is that the values of regulators are not given but should be computed at each time moment. Now $t$-fitness (3.1) is equal to

$$K_t(\mathbf{a})=\prod_{k=0}^t w_k(\mathbf{a})=\exp[\sum_{k=0}^t (\sum_{i=1}^n \varphi_i(\mathbf{a})u_i(S_i(k))+\sum_{j=1}^m \psi_j(\mathbf{a})v_j(H_j(k))].$$

Denote $\boldsymbol{\varphi}=(\varphi_1, \varphi_2, \ldots \varphi_n)$, $\boldsymbol{\psi}=(\psi_1, \ldots \psi_m)$ and let $p_t(\boldsymbol{\varphi},\boldsymbol{\psi})$ be the pdf of the random vector $(\boldsymbol{\varphi},\boldsymbol{\psi})$ at $t$ moment, i.e. $p_t(\mathbf{x},\mathbf{y}) = P_t\{\boldsymbol{\varphi}(\mathbf{a})=\mathbf{x}, \boldsymbol{\psi}(\mathbf{a})=\mathbf{y}\}$. Let $f(\mathbf{a})$ be a (measurable) function on $A$ and $\boldsymbol{\lambda}=(\lambda_1,\ldots \lambda_n)$, $\boldsymbol{\delta}=(\delta_1,\ldots \delta_m)$.

For given initial distribution $P_0(\mathbf{a})$, introduce the functional $M(f; \boldsymbol{\lambda}, \boldsymbol{\delta})$:

$$M(f; \boldsymbol{\lambda}, \boldsymbol{\delta})= \int_A f(\mathbf{a}) \exp[\sum_{i=1}^n \lambda_i\varphi_i(\mathbf{a})+\sum_{j=1}^m \delta_j\psi_j(\mathbf{a})]P_0(\mathbf{a})d\mathbf{a}. \qquad (4.2)$$

Denote $\mathbf{S}(t)=(u_1(S_1(t)),\ldots u_n(S_n(t)))$, $\mathbf{H}(t)=(v_1(H_1(t)),\ldots v_m(H_m(t)))$.

**Lemma 2**. $E_0[f K_t]= M(f; \sum_{k=0}^t \mathbf{S}(k), \sum_{k=0}^t \mathbf{H}(k))$.

**Theorem 2**. Let $P_0(\mathbf{a})$ be the initial pdf of the vector-parameter $\mathbf{a}$ for inhomogeneous map (4.1), and $M(f; \boldsymbol{\lambda}, \boldsymbol{\delta})$ be corresponding functional (4.2). Then

1) The total population size and the regulators can be computed with the help of the system of recurrence relations:

$$N_t = N_0 E_0[K_{t-1}]= N_0 M(1; \sum_{k=0}^{t-1} \mathbf{S}(k), \sum_{k=0}^{t-1} \mathbf{H}(k));$$

$$S_i(t) = N_0 E_0[s_i K_{t-1}]= N_0 M(s_i; \sum_{k=0}^{t-1} \mathbf{S}(k), \sum_{k=0}^{t-1} \mathbf{H}(k));$$

$$H_j(t) = E_0[h_j K_{t-1}]/E_0[K_{t-1}]= M(h_j; \sum_{k=0}^{t-1} \mathbf{S}(k), \sum_{k=0}^{t-1} \mathbf{H}(k))/M(1; \sum_{k=0}^{t-1} \mathbf{S}(k), \sum_{k=0}^{t-1} \mathbf{H}(k)).$$

2) The current pdf $P_t(\mathbf{a})$ is given by the formula
$P_t(\mathbf{a}) = P_0(\mathbf{a})K_{t-1}(\mathbf{a}) /E_0[K_{t-1}]$.

**5. The Price' equation and the FTNS**

Let $z_t(\mathbf{a})$ be a quantitative trait; the difference between mean values of the trait before and after selection (at $t$ time moment), $\Delta E_t[z]=E_{t+1}[z_{t+1}]-E_t[z_t]$, is known as the selection differential and is an important characteristic of selection. The covariance equation and the Price' equation show the connection between the selection differential and the fitness. The theory developed in s.3, s.4 allows us to overcome the problem of dynamical insufficiency of the Price' equations and the FTNS. Within the framework of models (2.1), (2.7) or (4.1), all statistical characteristics of interest could be computed effectively given the initial distribution.

**Proposition 1 (On the complete Price' equation).**
i) *Within the framework of master model* (2.1), (2.7) *all terms of the Price' equation,*
$\Delta_t E_t[z_t] = \{Cov_t[z_t w_t] + E_t[w_t \Delta_t z_t]\}/E_t[w_t]$ *can be computed explicitly*:
$E_t[z_t] = E_0[z_t K_{t-1}]/E_0[K_{t-1}]$,
$Cov_t[w_t, z_t]/E_t[w_t] = E_0[z_t K_t]/E_0[K_t] - E_0[z_t K_{t-1}]/E_0[K_t]$,
$E_t[w_t \Delta_t z_t]\}/E_t[w_t] = E_0[(z_{t+1} - z_t) K_t]/E_0[K_t]$ *and the selection differential is*
$\Delta_t E_t[z_t] = E_0[z_{t+1} K_t]/E_0[K_t] - E_0[z_t K_{t-1}]/E_0[K_{t-1}]$.

ii) *If the fitness is given by formula* (4.1), *then the t-fitness $K_t$ and mean values $E_0[K_t]$, $E_0[z_t K_{t-1}]$ can be computed recurrently with the help of Theorem 2.*

**Corollary (On the covariance equation).** *Let a trait $z(\mathbf{a})$ does not depend on time. Then all terms of the covariance equation $\Delta_t E_t[z] = Cov_t[zw_t]/E_t[w_t]$ can be computed explicitly, and the selection differential is $\Delta_t E_t[z] = E_0[zK_t]/E_0[K_t] - E_0[z K_{t-1}]/E_0[K_{t-1}]$.*

The FTNS is a particular case of the Price' equation under $z_t = w_t$.

**Proposition 2 (On the Fisher' FTNS).**
i) *In framework of master model* (2.1), (2.7) *all terms of the FTNS' equation*
$\Delta E_t[w_t] = \{Var_t[w_t] + E_t[w_t \Delta w_t]\}/E_t[w_t]$
*can be computed explicitly:*
$Var_t[w_t]/E_t[w_t] = E_0[w_t K_t]/E_0[K_t] - E_0[K_t]/E_0[K_{t-1}]$,
$E_t[w_t \Delta w_t]\}/E_t[w_t] = E_0[K_{t+1}]/E_0[K_t] - E_0[w_t K_t]/E_0[K_t]$,
*and the FTNS' equation is reduced to the difference equation*
$\Delta E_t[w_t] = E_0[K_{t+1}]/E_0[K_t] - E_0[K_t]/E_0[K_{t-1}]$.

ii) If $M(\lambda)$ is the mgf of the initial distribution of r.v. $\vartheta = (\varphi_1, \varphi_2, ... \varphi_n)$, then
$E_t[w_t] = E_0[K_t]/E_0[K_{t-1}] = M(\mathbf{G}(t))/M(\mathbf{G}(t-1))$
where $\mathbf{G}(t) = (G_1(t), ... G_n(t))$, $G_i(t) = g_1(s) + ... g_t(s)$.

iii) *In framework of model* (4.1) *the t-fitness $K_t$ and mean values $E_0[K_t]$ can be computed recurrently with the help of Theorem* 2.

## 6. Examples.

It is well known that non-linear population models with discrete time (maps) may demonstrate very complex and even counterintuitive behavior depending on the values of model parameters. The main dynamical regimes of the corresponding inhomogeneous models are crucially determined by the behaviors of the original homogeneous models, but have some essentially new interesting peculiarities due to "inner bifurcations", i.e., changing of parameters due to internal dynamics of the system.

### 6.1. Truncation selection

Let $z = z(\mathbf{a})$ be a quantitative trait and the individual fitness $w$ is completely defined by the value of this trait, $w = w(z)$. Then the population's evolution is governed by the equation $l(t+1, \mathbf{a}) = w(z(\mathbf{a})) l(t, \mathbf{a})$. The distribution of the trait may evolve in different ways depending on the initial pdf (see Remark at the end of s.3). An opposite situation occurs under the selection with strict truncation: $w(z) = C$, if $z \leq B = const$, and $w(z) = 0$ if $z > B$ (see [Crow & Kimura 1979], [Shnol&Kondrashov 1994]).
Let $p_B = P_0(z \leq B)$, then $E_0(w^t) = C^t p_B$. Hence, according to (3.5) $P_t(\mathbf{a}) = P_0(\mathbf{a})/p_B$, for $z(\mathbf{a}) \leq B$, and $P_t(\mathbf{a}) = 0$ for $z(\mathbf{a}) > B$. This means that $P_t(\mathbf{a})$ is equal to the conditional probability $P_0(\mathbf{a})$ under condition that $z \leq B$, i.e. $P_t(\mathbf{a}) = P(0, \mathbf{a})/P_0(z \leq B) \chi\{z(\mathbf{a}) \leq B\}$. Hence, the probability $P_t(\mathbf{a})$ does not change after the first selection step, $P_t(\mathbf{a}) = P_1(\mathbf{a})$ for all $t > 1$.
Selection differential is $\Delta E_0[z] = E_0[z \mid z \leq B] - E_0[z]$ at the first selection step and $\Delta E_t[z] = 0$ for any $t > 0$. Next, $E_t[w] = E_0[w^{t+1}]/E_0[w^t] = C$ and hence $N(t+1) = CN(t)$. Thus, the total size of the population increases (decreases) exponentially, $N(t) = N(0)C^t$, unless $C = 1$. A more realistic model should take into account the population-size regulation.

### 6.2. Logistic model

For a well known logistic map, $N_{t+1} = \lambda N_t (1 - N_t)$, $0 < \lambda < 4$ and $0 \leq N \leq 1$. Consider the inhomogeneous logistic model $l(t+1, a) = \lambda_0 l(t, a) a (1 - N_t)$, $\lambda_0 = const$, $a$ is the distributed parameter. Then $w_t(a) = \lambda_0 a(1-N_t)$. For this model, $E_t[w] = \lambda_0(1-N_t)E_0[a^{t+1}]/E_0[a^t]$, $N_{t+1} = N_t E_t[w]$. The model has a sense only if $0 < E_0[a^{t+1}]/E_0[a^t] < 4$.

Let $P_0(a)$ be the Beta-distribution in [0,1] with parameters $(\alpha,\beta)$. Then $E_t[w] = \lambda_0(t+\alpha)/(t+\alpha+\beta)(1-N_t)$, $N_{t+1} = N_t \lambda_0(t+\alpha)/(t+\alpha+\beta)(1-N_t)$. Choosing appropriate value of $\lambda_0 \leq 4$, we will observe (at $t \to \infty$) any possible behavior of the plain logistic model as the final dynamical behaviors of the inhomogeneous logistic model. In particular, at $\lambda_0 = 4$ (almost) all cycles of Feigenbaum's cascade appear in the course of time and realize as parts of the *single* trajectory, as a result of the "inner" bifurcations of the inhomogeneous logistic model. Figure 1 illustrates this assertion and shows also a very complex behavior of the mean fitness (in contrast with the plain FTNS).

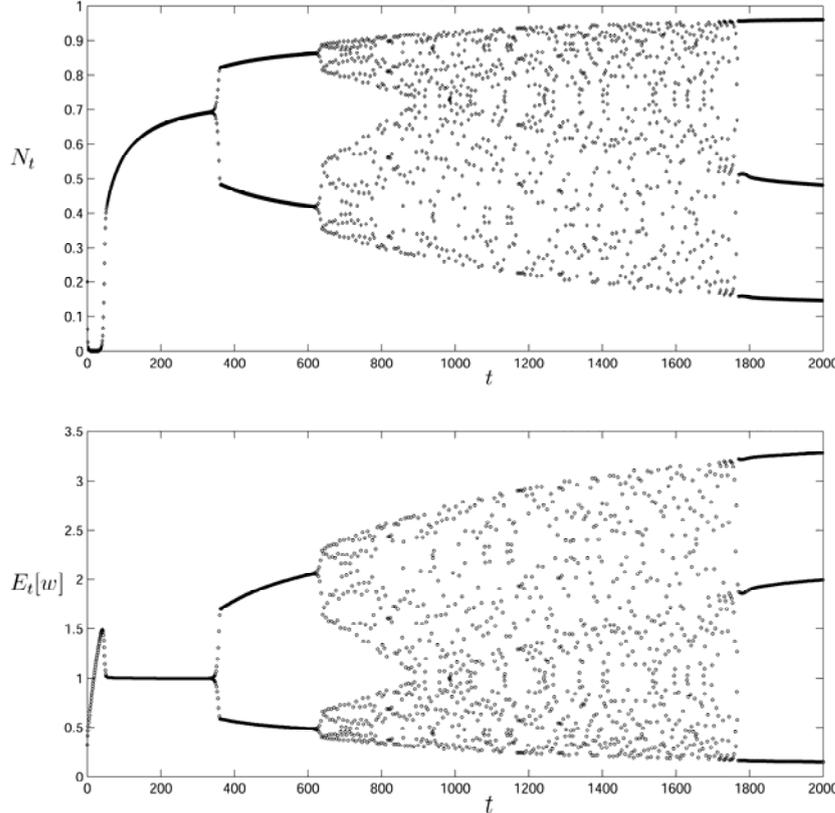

Fig. 1. The trajectory of total population size and mean fitness for inhomogeneous logistic model with Beta-distributed parameter ($\lambda_0=4$, $E_0[f]=0.1$, $Var_0[f]=0.005$).

### 6.3. Ricker' model

The classical *Ricker'* model $N_{t+1} = N_t \lambda \exp(-\beta N_t)$ where $\lambda$ and $\beta$ are positive parameters, takes into account the population size regulation of the reproduction rate. Consider the inhomogeneous population model with both distributed parameters $\alpha = \ln\lambda$ and $\beta$:

$l(t+1,\mathbf{a}) = l(t,\mathbf{a}) w(\mathbf{a}, N_t)$ where $w(\mathbf{a}, N_t) = \exp[\alpha(\mathbf{a}) - \beta(\mathbf{a})N_t]$.

Comparing with (4.1) we may put for this example $s(\mathbf{a})=1$, $u_1(x)=1$, $u_2(x)=-x$, so that $S(t)=N_t$, $u_1(S(t))=1$, $u_2(S(t))=-N_t$.

Let $M(\lambda_1, \lambda_2) = \int_A \exp(\lambda_1 \alpha(\mathbf{a}) + \lambda_2 \beta(\mathbf{a})) P_0(\mathbf{a}) d\mathbf{a}$

be the mgf of the initial distribution of r.v. $\alpha(\mathbf{a})$ and $\beta(\mathbf{a})$. Then

$$E_0[K_t] = M(t+1, -\sum_{k=0}^{t} N_k). \quad (6.1)$$

Applying Theorem 2 we obtain

$$N_t = N_0 E_0[K_{t-1}] = N_0 M(t, -\sum_{k=0}^{t-1} N_k); \quad (6.2)$$

$$P_t(\mathbf{a}) = P_0(\mathbf{a}) \exp[t\alpha(\mathbf{a}) - \beta(\mathbf{a})\sum_{k=0}^{t-1} N_k]/ M(t, -\sum_{k=0}^{t-1} N_k). \quad (6.3)$$

These formulas completely solve the inhomogeneous Ricker' model.

The selection differential for the model is $\Delta E_t[w_t] = E_0[K_{t+1}]/E_0[K_t] - E_0[K_t]/E_0[K_{t-1}]$ where $E_0[K_t]$ for given initial distributions can be computed recurrently by formula (6.1).

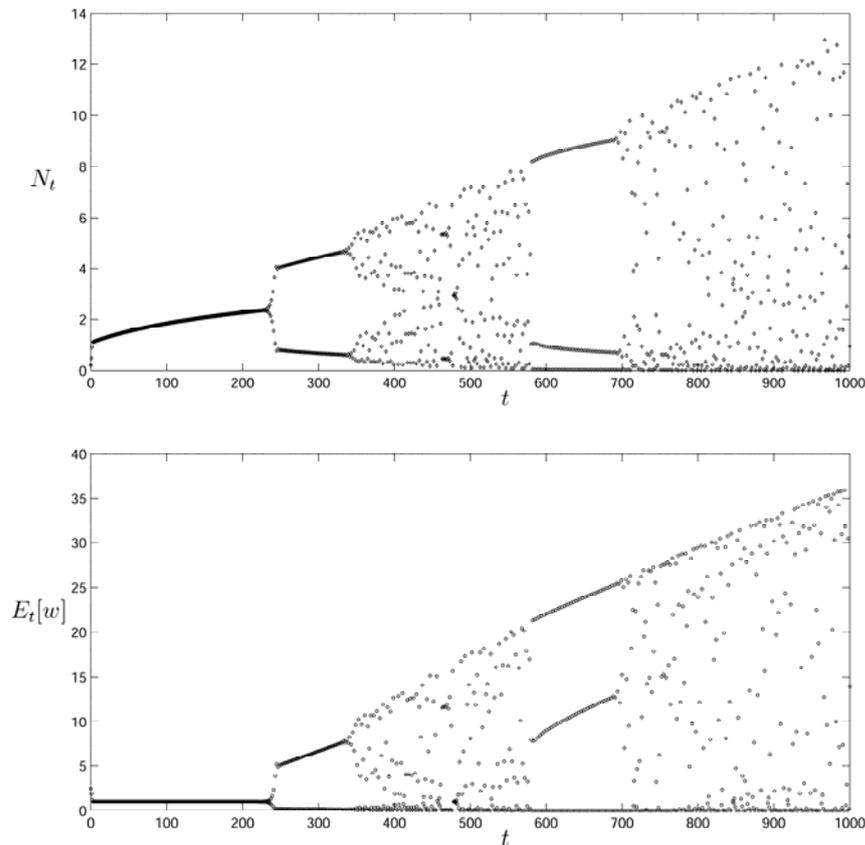

Figure 2. The trajectory of total population size and mean fitness for inhomogeneous Ricker' model with $\Gamma$-distributed parameters $a$ ($E_0[a]$ =3, $Var_0[a]$=0.1).

### 6.4. Selection in a Natural Rotifer Community

The mathematical model of zooplankton populations was suggested in [Snell 1998] and studied systematically in [Berezovskaya 2005]. The model depends on the parameters $a$ characterizing the environment quality and $\gamma$ that is the species-specific parameter. Let us consider the model of a community that consists from different rotifer populations; individuals inside the populations may have different reproduction capacities under constant toxicant exposure. The model is of the form $l(t+1,\mathbf{a})=l(t,\mathbf{a})w_t(\mathbf{a})$ where $w_t(\mathbf{a}) =\exp[-a +1/N_t-\gamma/N_t^2]$ and $\mathbf{a}=(a,\gamma)$. Let $M(\lambda_1,\lambda_2)$ be the mgf of the initial distribution of $a$ and $\gamma$. Then $N_t= N_0\exp[\sum_{k=0}^{t-1} 1/N_k]M(-t,-\sum_{k=0}^{t-1} 1/N_k^2)$.

The trajectory $N_t$ has a very complex transition regime from the initial to the final behavior, see Figure 3 (for both Gamma- distributed independent parameters $a,\gamma$).

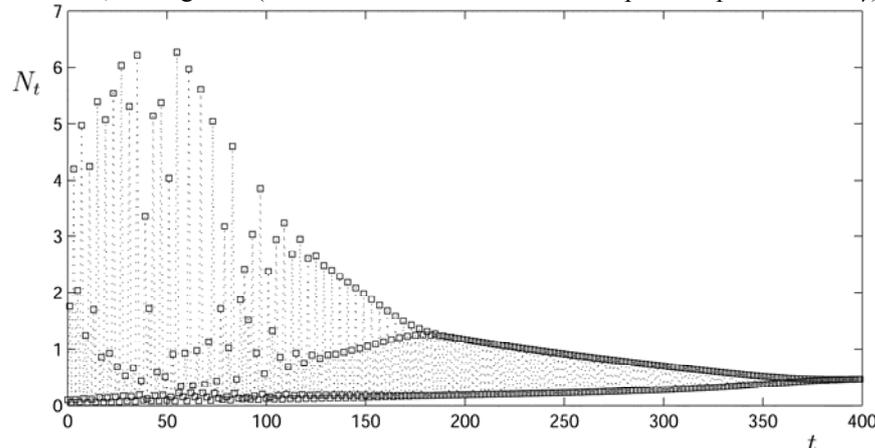

Figure 3. The trajectory of total population size for inhomogeneous rotifer' model with $\Gamma$-distributed parameters $a$ ($E_0[a]$ =4, $Var_0[a]$=0.01) and $\gamma$ ($E_0[\gamma]$ =0.06, $Var_0[a]$=0.03, $b$=0.03).

**Conclusion**

Mathematical theory of selection systems is developed here for a wide class of dynamical models with discrete time. The Price' equation and the Fisher' Fundamental theorem of natural selection together with the Haldane' principle are well known general results of the mathematical selection theory. The Price' equation is a mathematical identity and so is not dynamically sufficient.

We show that the problem of dynamical insufficiency for the Price' equations and for the FTNS can be resolved to these equations are analyzed on the basis of the entire initial distribution within the framework of inhomogeneous dynamical models. For these models, the current distribution and, accordingly, all statistical characteristics of interest could be computed effectively for any time instance.

**Acknowledgement.** Author thanks Dr. E. Koonin and Dr. A. Novozhilov for valuable discussions and help in preparation of the manuscript.